\documentstyle[12pt,a4,epsfig]{article}
\newlength{\abstwidth}
\begin{document}
\newcommand{\hi }{{\footnotesize HIJING}}
\newcommand{\hij}{{\footnotesize HIJING }}
\newcommand{\qgp}{{\footnotesize QGP}}
\newcommand{\qgpp}{{\footnotesize QGP }}
\newcommand{\qcd}{{\footnotesize QCD}}
\newcommand{\qcdd}{{\footnotesize QCD }}
\newcommand{\pqcd}{{\footnotesize pQCD }}
\newcommand{\lhc}{{\footnotesize LHC }}
\newcommand{\rhic}{{\footnotesize RHIC }}
\newcommand{\nudy}{$\nu_{dyn}$}
\newcommand{\nch}{$\langle N_{ch} \rangle$}
\newcommand{\nchh}{$\langle N_{ch} \rangle$ }
\newcommand{\dnch}{$\langle dN_{ch}/d\eta \rangle$}
\newcommand{\dnchh}{$\langle dN_{ch}/d\eta \rangle$ }
\newcommand{\np}{$\langle N_{part} \rangle$}
\newcommand{\npp}{$\langle N_{part} \rangle$ }
\newcommand{\myRed}[1]{\textcolor{red}{#1}}

\thispagestyle{empty}
\begin{center}
{\Large {\bf Event-by-Event Particle Ratio Fluctuations at LHC Energies }}\\[5mm]

{\bf Shaista Khan\footnote{email: Shaista.Khan@cern.ch \\ \noindent Authors declare that there is no conflict of interest}, Bushra Ali, Anuj Chandra and Shakeel Ahmad  }\\
{\it  Department of Physics, Aligarh Muslim University\\[-2mm] Aligarh - 202002 INDIA}
\end{center}

\begin{center}
{\bf Abstract}\\[2ex]
\begin{minipage}{\abstwidth} A Monte Carlo study of identified particle ratio fluctuations at {\footnotesize LHC} energies is carried out in the frame work of \hij model using the fluctuation variable $\nu_{dyn}$. The simulated events for Pb-Pb collisions at $\sqrt{s}_{NN}$ = 2.76 and 5.02 TeV and Xe-Xe collisions at $\sqrt{s}_{NN}$ = 5.44 TeV are analyzed. From this study, it is observed that the values of $[\pi,K]$, $[p,K]$ and $[\pi,p]$ follow the similar trends of energy dependence as observed in the most central collision data by {\footnotesize NA49}, {\footnotesize STAR} and {\footnotesize ALICE} experiments. It is also observed that $\nu_{dyn}$ for all the three combinations of particles for semi-central and central collisions, the model predicted values of $\nu_{dyn}[A,B]$ for Pb-Pb collisions at $\sqrt{s}_{NN}$ = 2.76 TeV agree fairly well with those observed in {\footnotesize ALICE} experiment. For peripheral collisions, however, the model predicted values of $\nu_{dyn}[\pi,K]$ are somewhat smaller, whereas for $[p,K]$ and $[\pi,p]$ it predicts larger values as compared to the corresponding experimental values. The possible reasons for the observed differences are discussed. The $\nu_{dyn}$ values scaled with charged particle density when plotted against $\langle$N$_{part}$$\rangle$, exhibit a flat behaviour, as expected from the independent particle emission sources. For $[p,K]$ and $[\pi,p]$ combinations, a departure from the flat trend is, however, observed in central collisions in the case of low p$_{T}$ window when effect of jet quenching or resonances are considered. Furthermore, the study of $\nu_{dyn}[A,B]$ dependence on particle density for various collision systems (including proton-proton collisions) suggests that at {\footnotesize LHC} energies 
$\nu_{dyn}$ values for a given particle pair is simply a function of charged particle density, irrespective of system size, beam energy and collision centrality.\\
	
 PACS numbers: 25.75--q, 25.75.Gz \\[3ex]
\end{minipage}
\end{center}
\noindent KEY-WORDS: Event-by-event particle ratio fluctuations, Relativistic heavy-ion collisions.

\newpage
\noindent {\bf Introduction}\\
\noindent Fluctuations associated to a physical quantity measured in an experiment, in general, depend on the property of the system and are expected to provide useful clue about the nature of the system under study \cite{bib:1,bib:2,bib:3}. As regards the heavy-ion (AA) collisions, the system created is assumed to be a hot and dense fireball of hot partonic and (or) hadronic matter \cite{bib:1,bib:2}. One of the main aims of studying AA collisions at relativistic energies is to search for the existence of partonic matter in the early stage of the created fireball. Fluctuations associated to a thermal system are supposed to be related to various susceptibilities \cite{bib:1,bib:2,bib:4} and would serve as an indicator of the possible phase transition. Moreover, the presence of large event-by-event (ebe) fluctuations, if observed, might be signal for the presence of distinct classes of events, one with and one without \qgpp formation \cite{bib:5,bib:6,bib:7}. Therefore, the search for the phase transition from hadronic matter to  \qgpp still remains a topic of interest of high energy physicists \cite{bib:8,bib:9,bib:10}. Correlations and ebe fluctuations of dynamical nature are believed to be associated with the critical phenomena of phase transition and their studies would lead to the local and global differences between the events produced under similar initial conditions \cite{bib:11}.\\

\noindent ebe fluctuations in hadronic and heavy-ion collisions have been investigated at widely different energies using several different approaches, for example, normalized factorial moments \cite{bib:12,bib:13,bib:14,bib:15}, multifractals \cite{bib:16,bib:17}, k-order rapidity spacing \cite{bib:18,bib:19,bib:20}, erraticity \cite{bib:21,bib:22,bib:23}, intensive and strongly intensive quantities (defined in term of multiplicity, transverse momentum, p$_{T}$, etc.) \cite{bib:24,bib:25,bib:26}. Furthermore, ebe fluctuations in conserved quantities like strangeness, baryon number and electric charge have emerged as new tools to estimate the degree of equilibration and criticality of the measured systems \cite{bib:27,bib:28,bib:29,bib:30,bib:31}. The dynamical net charge fluctuations have been investigated by {\footnotesize STAR} and {\footnotesize ALICE} experiments \cite{bib:28,bib:29} in terms of variable $\nu_{dyn}$ \cite{bib:32}, which is an excellent probe because of its robustness against detector efficiency losses \cite{bib:29}. The other measures of the net charge fluctuations, like the variance of charge V(Q), variance of charge ratio V(R), and the D-measure \cite{bib:29,bib:31,bib:33,bib:34} prone to the measurement conditions \cite{bib:32,bib:35}.\\

\noindent It has, however, been pointed out \cite{bib:36} that large systematic uncertainties, like volume fluctuations due to impact parameter variations are associated in such measurements, while the multiplicity ratio fluctuations are sensitive to the density fluctuations instead of volume fluctuations \cite{bib:jeon1999}. Thus, the variable $\nu_{dyn}$, defined by considering the particle species pair, rather than defining it in terms of combinations of like and unlike charges, has been used as a tool to probe the properties of \qgp \cite{bib:33,bib:37}. Since it is speculated that the phase transition, if occurs, would result in increase and divergence of fluctuations and could be related to ebe fluctuations of a suitably chosen observable. An enhanced fluctuations in the particle ratio is expected during a phase transition at critical point (CP). For example, $[\pi,K]$, $[p,K]$ and $[\pi,p]$ fluctuations could be related to baryon number fluctuations, strangeness fluctuations and baryon-strangeness correlations \cite{bib:37,bib:38}. \\

\noindent Particle ratio fluctuations in AA collisions have been addressed in a number of studies, e.g., {\footnotesize NA49} experiment in Pb-Pb collisions at E$_{lab}$ = 20-158 A GeV \cite{bib:39}, {\footnotesize STAR} experiment in Au-Au collisions at $\sqrt{s}_{NN}$ = 7.7 to 200 GeV \cite{bib:40}, Cu-Cu collisions at $\sqrt{s}_{NN}$ = 22.4, 62.4 and 200 GeV \cite{bib:41} and several others \cite{bib:33,bib:36,bib:37}. At {\footnotesize LHC} energies, the particle ratio fluctuations has been investigated by {\footnotesize ALICE} experiment at $\sqrt{s}_{NN}$ = 2.76 TeV only \cite{bib:42,bib:mesut,bib:43}. It has been reported \cite{bib:42} that $\nu_{dyn}$ for $[\pi,K]$ and $[p,K]$ combinations acquires positive values irrespective of the centrality class, whereas, for $[\pi,p]$ combination, the variable changes sign from positive to negative toward more peripheral collisions, indicating the difference in the production mechanisms involved of these pairs. The observed trend of energy dependence of $\nu_{dyn}$ with beam energy \cite{bib:42} suggests that the production dynamics changes significantly from that reported at lower energies. It has also been pointed out \cite{bib:42} that further investigations involving fluctuations with charge and species specific pairs be carried out to characterize the production dynamics and understand the observed sign changes. It was, therefore, considered to undertake the study of particle ratio fluctuations by analysing the data on Pb-Pb collisions at $\sqrt{s}_{NN}$ = 2.76 and 5.02 TeV and Xe-Xe collisions at $\sqrt{s}_{NN}$ = 5.44 TeV in the framework of \hij model. Using the \hij the effect of jet quenching and resonance production can also be looked into. \\
 
 \noindent {\bf Formalism} \\
 \noindent The particle ratio fluctuations may be studied in terms of the yields of the ratio of particle types A and B. The particle ratio A/B is estimated by counting the particle types A and B produced in each event. Using the relative widths of the particle ratio distributions of the data and the corresponding mixed events the observable $\sigma_{dyn}$ is defined as \cite{bib:37,bib:42a}:\\
 \begin{eqnarray} 
	\sigma_{dyn} = sgn(\sigma^2_{data} - \sigma^2_{mixed})\sqrt{|\sigma^2_{data} - \sigma^2_{mixed}|}
\end{eqnarray}
\\
\noindent where $\sigma_{data}$ and $\sigma_{mixed}$ respectively denote the relative widths (standard deviation/mean) of the ratio A/B for the data and mixed events. Yet another variable $\nu_{dyn}$, which is commonly accepted for studying the particle ratio fluctuations has been proposed \cite{bib:32}. $\nu_{dyn}[A,B]$ quantifies the deviation of the fluctuations in the number of particle species A and B from that expected from Poissonian statistics \cite{bib:42a}. This variable does not involve particle ratios directly but is related to $\sigma_{dyn}$ as $\sigma_{dyn}^{2}$ $\approx$ $\nu_{dyn}[A,B]$ \cite{bib:41,bib:42a}. \\ 
 
\noindent The $\nu_{dyn}[A,B]$ is defined as \cite{bib:37,bib:42,bib:43,bib:42a}:\\
  \begin{eqnarray} 
 	 \nu_{dyn}[A,B] = \frac{\langle N_A(N_A - 1)\rangle}{\langle N_A\rangle^2} + \frac{\langle N_B(N_B - 1)\rangle}{\langle N_B\rangle^2} -2\frac{\langle N_AN_B\rangle}{\langle N_A\rangle\langle N_B\rangle}  	 
 \end{eqnarray}
 \\
\noindent where $\langle N_{A}\rangle$ and $\langle N_{B}\rangle$  respectively denote the event multiplicities of particle types A and B within the given kinematical limits, while the quantities within $\langle...\rangle$ represent their mean values. It should be mentioned here that the particle type A or B includes the particle and its anti-particle.
 $\nu_{dyn}[A,B]$ basically contrasts the relative strengths of fluctuations of particle type A and B to the relative strength of correlation between the types A,B. It may be noted that $\nu_{dyn}[A,B]$ should be zero if particles A and B are produced in statistically independent way \cite{bib:32,bib:35,bib:42}. In practice, however, a non-zero value of $\nu_{dyn}[A,B]$ is expected because produced particles are partially correlated through the production of resonances, string fragmentation, jet fragmentation and (or) other mechanisms \cite{bib:29}. A negative value of $\nu_{dyn}[A,B]$ indicates a correlation, whereas positive value would indicate the presence of anti-correlation between particle types A and B. The indices A, B are taken as particle pair combinations, such as $[\pi,K]$, $[p,K]$ and $[\pi,p]$ in the present work to construct the $\nu_{dyn}$. \\
	 
\noindent {\bf The HIJING model} \\
\noindent The Monte Carlo model \hij (Heavy-Ion Jet Interaction Generator) was developed to study the role of minijets and particle production in proton-proton (pp), proton-nucleus (pA) and nucleus-nucleus (AA) collisions in a wide range of energies from 5 to 2000 GeV \cite{bib:Gyulassy_1994,bib:Wang_1991}. The \hij model is commonly used in high energy heavy-ion collisions for providing the baseline to compare the simulation results with the experimental data. The main feature of \hij model is based on \pqcd (perturbative \qcd) approach considering that the multiple minijet partons produced in collisions are transformed into string fragmentation which, in turn, decays into hadrons. The \pqcd process is implemented in \hij using {\footnotesize PYTHIA} \cite{bib:Sjostrand_2019,bib:Bengtsson} model for hadronic collisions. The cross-section in \pqcd for hard parton scattering is determined using leading order to simulate the higher order corrections. The eikonal formalism is embedded to calculate the number of minijets per inelastic nucleon-nucleon collisions. The soft contributions are modeled by diquark-quark strings with gluon kinks along with the line of the {\footnotesize Lund FRITIOF} and {\footnotesize DPM} (Dual Parton Model) \cite{bib:Wang_1991,bib:Anderson,bib:Nilson_1987,bib:Capella,bib:Ranft1,bib:Ranft2}. Besides this, the basic property of \hij model is that it considers the nucleus-nucleus collisions as a superposition of nucleon-nucleon collisions. However, the mechanism for final state interactions among the low p$_{T}$ particles is not included in the \hij model. Due to which the phenomena such as collectivity and equilibrium can not be addressed. Therefore, \hij is mainly designed to explore the range of possible initial conditions that may occur in high energy heavy-ion collisions.\\

\noindent Furthermore, \hij also takes into account other important physics processes like jet quenching \cite{bib:Gulassy_1990}, multiple scattering and nuclear shadowing to study the nuclear effects \cite{bib:Wang_1991}. To study the dependence of moderate and high p$_{T}$ observables on an assumed energy loss of partons traversing the produced dense matter, a jet quenching approach is incorporated in the \hij model \cite{bib:Wang_1991}.\\

\noindent In high energy heavy-ion collisions, the interaction of high p$_{T}$ jets in the produced transient dense medium is treated as one of the signals of phase transition \cite{bib:Gulassy_1990}. Therefore, the rapid-variation of $\mu_{D}$ (Debye Screening) near the phase transition point could lead to a variation of jet quenching phenomenon, that could be used as a diagnostic tool of the \qgpp phase transition \cite{bib:Wang_1991}. Furthermore, resonances play an important role in studying the net-charge fluctuations. Resonances have short lifetime, and subsequently decay into stable hadrons. This would affect the final hadron yields and their number fluctuations \cite{bib:resonance1}. Resonance decay kinematics influences charge fluctuations in two different ways. It dilutes the effect of global charge conservation if only one of the decay products falls into the acceptance window. However, if both decay products lie within the acceptance cone, mean charged particle multiplicity will increase but the net charge does not change \cite{bib:resonance2}. Hadron production in \hij involves a cocktail of resonances that may also give a rough estimate of the strength of correlations between charged and neutral kaons \cite{bib:resonance3}. Present study is an attempt to explore the effect of fluctuations in understanding the dissipative properties of a color defined medium using the jet quenching, resonance production and jet/minijet contributions incorporated in \hij model \cite{bib:Liu_2003}. It was found that \cite{bib:Singh_2013, bib:Deng_2011,bib:Armesto_2000,bib:Armesto_2005} the \hij predicted values of charged particle density when jet quenching and contributions from resonance decays are switched off, are consistent with the ones observed in Au-Au collisions at 200 GeV per nucleon and Pb-Pb at 2.76 and 5.5 TeV per nucleon.  \\

\noindent {\bf Results and Discussion} \\ 
\noindent MC events corresponding to Pb-Pb at $\sqrt{s}_{NN}$ = 2.76, 5.02 and Xe-Xe collisions at $\sqrt{s}_{NN}$ = 5.44 TeV are generated using the \hi-1.37 \cite{bib:Gyulassy_1994,bib:Wang_1991}. Events are simulated by running the code in three different modes; (i) \hij default, i.e Resonance (Res) off, Jet Quenching (JQ) off (ii) Res-on JQ-off and (iii) Res-off JQ-on. The number of events simulated in each of these modes are listed in Table~1. The analysis is carried out by considering only those charged particles which have pseudorapidity ($\eta$) and transverse momentum (p$_{T}$) in the range, $|\eta|$ $<$ 0.8 and $0.2 < p_{T} < 1.5$ GeV/c respectively. {\footnotesize ALICE} experiment has also used same $\eta$-cut but instead of p$_{T}$ they have considered the charged particles with momentum, $0.2 < p < 1.5$ GeV/c. It may be mentioned here that for the $\eta$ range considered in {\footnotesize ALICE} experiment and also in the present study, for $p_{T} < 5.0$ GeV/c, $p_{T} \simeq 0.9p$. In order to examine the effect of jet quenching, a higher p$_{T}$ range, $0.2 < p_{T} < 5.0$ GeV/c, is also considered where this effect is expected to be more visible. The centrality of an event is estimated by applying {\footnotesize VZERO-A} and {\footnotesize VZERO-C} detector $\eta$ cuts of the {\footnotesize ALICE} experiment \cite{bib:63,bib:64,bib:65}, i.e., by considering the charged particles which have their $\eta$ values in the range, $2.8 <\eta <5.1$ or $-3.7 < \eta < -1.7$. For this multiplicity distributions of charged particles having their $\eta$ values within these limits are examined and quantiled to fix the minimum and maximum limits for a centrality class. \\

\noindent The values of mean number of participating nucleons, \npp  and mean charged particle density, \dnchh for different centrality classes are listed in Tables~2-4. Variations of \dnchh with \npp for these events are plotted in Fig.1. The values of \npp and \dnchh reported earlier \cite{bib:63,bib:64,bib:65} are also given in these tables and displayed in the figure. It is interesting to note from Tables~2-4 and Fig.1 that \hi-default predicts somewhat higher values of \npp and \dnchh for various centrality classes as compared to those observed in experiments \cite{bib:63,bib:64}. It may also be noted from the tables and the figure that the values of \dnchh are higher when jet quenching is turned on, which might be due to the enhanced production of low p$_{T}$ particles. This may be understood as when a partonic jet is quenched in the dense medium, it would fragment into large number of partons which, in turn, result in the production of low p$_{T}$ charged particles \cite{bib:66}. It may also be noted that the effect of jet quenching is rather more pronounced in central collisions, as compared to that in peripheral collisions. Enhancement in the \dnchh values due to resonances may also be seen in the figure. \\

\noindent Variations of mean multiplicities of charged pions, kaons, protons and anti-protons with \npp are shown in Fig.2. It is observed that mean multiplicities of \(\pi^\pm\), \(K^\pm\) and \(p\overline{p}\) increase with increasing \npp in almost identical fashion. It is also noted that the contributions to the particle multiplicities due to the jet quenching and resonance decays are maximum in most central collisions, which gradually decrease with \npp and tend to vanish for \npp values corresponding to centrality $\sim$ 50$\%$ and above. The reasons for the enhancement in particle multiplicities have been discussed in the previous section.\\

\noindent The values of $\nu_{dyn}$ for the combinations of particles [$\pi$,K], [p,K] and [$\pi$,p]  are calculated for various centrality classes using Eq.1. Variations of $\nu_{dyn}$ for these species of particles for 0-5$\%$ central collisions with beam energy are shown in Fig.3. Values of $\nu_{dyn}$ for these combinations of particles, reported by {\footnotesize NA49} \cite{bib:39}, {\footnotesize STAR} \cite{bib:40} and {\footnotesize ALICE} experiment \cite{bib:42} are also presented in the figure. Kinematical ranges used in these experiments are mentioned in the figure.  \\

\noindent The statistical errors associated to $\nu_{dyn}$ are too small to be visible in the figure. These errors are determined using the sub-sample method \cite{bib:39}: The data set is divided into 30 sub-samples and the values $\nu_{dyn}[A,B]_{i}$ are calculated for each sub-sample independently. Using these values of $\nu_{dyn}$, the mean and dispersion are estimated as; \\
\begin{eqnarray}
         <\nu_{dyn}[A,B]> = \frac{1}{n}\Sigma\nu_{dyn}[A,B]_i
\end{eqnarray}

\begin{eqnarray}  
	 \sigma_{\nu_{dyn}} = \sqrt{ \frac{\Sigma(\nu_{dyn}[A,B]_i - <\nu_{dyn}[A,B]>)^{2}}{n-1}  } 
\end{eqnarray}

The statistical error associated is then calculated as;

\begin{eqnarray}  
	(Error)_{stat} = \frac{\sigma_{\nu_{dyn}}}{\sqrt{n}} 
\end{eqnarray}

The following observations may be made from the figure:
\begin{itemize}
	\item $\nu_{dyn}[\pi,K]$ measured by {\footnotesize STAR} and {\footnotesize ALICE} experiments \cite{bib:40,bib:42} acquire positive and nearly energy independent values from $\sqrt{s}_{NN}$ = 7.7 GeV to 2.76 TeV. The \hij estimated values for $\sqrt{s}_{NN}$ = 2.76 TeV Pb-Pb collisions in the present study are observed to be close to that reported by {\footnotesize ALICE} experiment. The values of $\nu_{dyn}[\pi,K]$ for 0--3.5$\%$ Pb-Pb collisions also match with the {\footnotesize STAR} findings at $\sqrt{s}_{NN}$ = 11.5 and 19.6 GeV while below $\sqrt{s}_{NN}$ = 11.5 GeV an increasing trend in $\nu_{dyn}[\pi,K]$ is seen with decreasing beam energy. Such a difference in $\nu_{dyn}[\pi,K]$ values observed in {\footnotesize NA49} \cite{bib:39} and {\footnotesize STAR} \cite{bib:40} experiments has been argued to be due to the difference in measurement methods adopted in the two experiments. The observed positive values of $\nu_{dyn}[\pi,K]$ in experiments from $\sqrt{s}_{NN}$ = 7.7 GeV to 2.76 TeV as well as predicted by {\footnotesize UrQMD}, {\footnotesize HSD} \cite{bib:40} and \hij in the present study are either due to the dominance of variance of K and $\pi$ or because of the presence of an anti-correlation ($\langle$N$_{\pi}$N$_{K}$ $\rangle$ $<$ 0) between the K and $\pi$.\\
	
	\item $\nu_{dyn}[p,K]$ values, as reported by {\footnotesize STAR} \cite{bib:40} and {\footnotesize ALICE} \cite{bib:42}, may be observed to show an increasing trend with beam energy. At $\sqrt{s}_{NN}$=7.7 GeV the value is maximum negative, approaches to zero at $\sqrt{s}_{NN}$ = 200 GeV and becomes positive for Pb-Pb collisions at $\sqrt{s}_{NN}$ = 2.76 TeV. This indicates that the correlation between kaons and protons decreases with increasing incident energy. The \hij values observed at $\sqrt{s}_{NN}$ = 2.76 TeV are close to the experimental results. The \hij data points for $\sqrt{s}_{NN}$ = 5.02 TeV Pb-Pb collisions and $\sqrt{s}_{NN}$ = 5.44 TeV Xe-Xe collisions tend to follow the trend shown by the data, if extrapolated upto these energies. The higher and positive values of $\nu_{dyn}[p,K]$ observed by {\footnotesize NA49} experiment \cite{bib:39} for 0-3.5$\%$ central Pb-Pb collisions might be due to different detector acceptance of {\footnotesize NA49} and {\footnotesize STAR} experiments; detection of particle pairs resulting from the resonance decays are affected by the limited detector acceptance \cite{bib:40}. Studies involving second-order off-diagonal cumulants in the energy range, $\sqrt{s}_{NN}$ = 7.7 to 200 GeV carried out by {\footnotesize STAR} experiment \cite{bib:STARcummulants} show that the correlations between net proton and net kaon multiplicity distributions is negative at $\sqrt{s}_{NN}$ = 200 GeV. It increases with decreasing beam energy, changes sign at $\sqrt{s}_{NN}$ = 19.6 GeV and is maximum at $\sqrt{s}_{NN}$ = 7.7 GeV. The possible reason for the positive correlation between net proton and net kaons might be due to the associated production:$pp \rightarrow p \Lambda(1115)K^{+}$ \cite{bib:netproton}, which would give events having higher net proton to be associated to higher net kaons at lower energies. It has been argued that the negative correlation between $p$ and $k$ is expected to arise from \qgpp phase, where T-$\mu_{B}$ dependence is weak. Although the model calculations based on non-thermal ({\footnotesize UrQMD}) and thermal ({\footnotesize HRG}) production of hadrons do not agree with the experimental results, but it is expected that such a data-model comparison using the data with improved tracking capabilities and enhanced acceptance will help to understand the baryon-strangeness correlations which is predicted to have different T-$\mu_{B}$ dependence in hadronic and \qgpp phases.\\
	
	\item $\nu_{dyn}[\pi,p]$ values from {\footnotesize STAR} \cite{bib:40} and {\footnotesize ALICE} \cite{bib:42} experiments exhibit almost similar trend of energy dependence as that in the case of $\nu_{dyn}[p,K]$. $\nu_{dyn}[\pi,p]$ values for Pb-Pb collisions at $\sqrt{s}_{NN}$ = 2.76 TeV, observed from \hij model, when resonance and jet quenching are switched off, are close to the reported experimental values. It may also be noted that the $\nu_{dyn}[\pi,p]$ values obtained with resonance turned as `on' are somewhat higher for Pb-Pb collisions at both $\sqrt{s}_{NN}$ = 2.76 and 5.02 TeV. This might be because of relative dominance of $\Delta$ resonance production predicted by the model which give rise to pion-proton correlations as compared to uncorrelated p$\bar{p}$ pair production \cite{bib:40}. The reduction in $\nu_{dyn}[\pi,p]$ values observed in experiments has been argued to be due to increasing rate of pair production as compared to the rate of $\Delta$ resonance production with increasing beam energy.	
\end{itemize}


\noindent Variation of $\nu_{dyn}$ with $\langle$N$_{part}$$\rangle$ for the three combinations of particle species are shown in Fig.4. It is observed that $\nu_{dyn}$ is maximum for the smallest value of $\langle$N$_{part}$$\rangle$, i.e., for peripheral collisions. It decreases quickly as $\langle$N$_{part}$$\rangle$ becomes larger and thereafter acquire nearly constant positive values for $\langle$N$_{part}$$\rangle$ $\ge$ 100. In order to examine the effect of jet quenching, a parallel analysis of the data considering the p$_{T}$ range 0.2 $<$ p$_{T}$ $<$ 5.0 GeV/c is also carried out, because this effect is expected to be more visible on higher p$_{T}$ range. The values of $\nu_{dyn}$ for this p$_{T}$ range are plotted against $\langle$N$_{part}$$\rangle$ in Fig.5. The values of $\nu_{dyn}[A,B]$ for Xe-Xe collisions at $\sqrt{s}_{NN}$ = 5.44 TeV are noticed to be larger as compared to those observed for Pb-Pb collisions at $\sqrt{s}_{NN}$ = 2.76 and 5.02 TeV in the region of low $\langle$N$_{part}$$\rangle$. This indicates the presence of rather stronger anti-correlation in peripheral collisions in the case of smaller systems. The effect of jet quenching and resonance decays are also seems to be absent except for very peripheral collisions. For [$\pi$,K] combination \hij predicts slightly smaller values of $\nu_{dyn}$ for semi-central and peripheral collisions while for [p,K] pair the model overestimates $\nu_{dyn}$ as compared to those obtained from the data \cite{bib:42}. For [$\pi$,p] combinations, experimental results for Pb-Pb collisions at $\sqrt{s}_{NN}$ = 2.76 TeV show that the values of $\nu_{dyn}$ decrease with increasing centrality and became more negative for collisions centrality $>$ 40$\%$. It may also be noted from Figs.4 and 5 that $\nu_{dyn}$ values for various combinations of particle pairs are similar to those obtained with p$_{T}$ cut, 0.2 $<$ p$_{T}$ $<$ 1.5 GeV/c.\\   

\noindent Although $\nu_{dyn}$ is robust against detector efficiency losses, yet it has some intrinsic multiplicity dependence \cite{bib:42,bib:67}. In order to reduce the effect of multiplicities, the $\nu_{dyn}$ values for all three combinations of particles are scaled by mean charged particle density, \dnch. This removes the $1/N_{ch}$ dependence of $\nu_{dyn}$ \cite{bib:32,bib:41}. The scaled values of $\nu_{dyn}[A,B]$ with \dnchh are plotted as a function of \npp for $|\eta|$ $<$ 0.8 and p$_{T}$  = 0.2 to 1.5 and 0.2 to 5.0 GeV/c in Figs.6 and 7. Experimental results for Pb-Pb collisions at $\sqrt{s}_{NN}$ = 2.76 TeV for $|\eta|$ $<$ 0.8 and p = 0.2 to 1.5 GeV/c are also shown in Fig.6. It may be observed from Figs.6 and 7 that;
\begin{enumerate}
	\item $\nu_{dyn}[\pi,K]$ scaled values are positive and nearly independent of collision centrality and p$_{T}$ cuts applied. These values are close to those reported by the {\footnotesize ALICE} experiment \cite{bib:42} for the same $\eta$ cut and p = 0.2 to 1.5 GeV/c. The effect of jet quenching and resonance decay are, however, noticed to be absent.
	
	\item In case of \hij default, scaled values of $\nu_{dyn}[p,K]$ and $\nu_{dyn}[\pi,p]$ for p$_{T}$ = 0.2 to 1.5 GeV/c are observed to acquire nearly constant values against centrality and beam energy. These values are, however larger than those observed in {\footnotesize ALICE} experiment. It is also seen in the figure that the values are higher for central collisions when resonance production is switched on, but on increasing the p$_{T}$ range, i.e., 0.2 to 5.0 GeV/c, the effect of resonance vanishes. 
\end{enumerate}

\noindent The scaled $\nu_{dyn}$ values obtained by {\footnotesize ALICE} collaboration using \hij model may be noticed to be positive nearly constant against dN$_{ch}$/d$\eta$ \cite{bib:42,bib:mesut} for all centrality classes and for all the three combinations of particle species. These findings are found to be in good agreement with the values obtained in the present study. This suggests that although \hij implements global conservation laws yet it does not exhibit any non monotonic behaviour as a function of centrality. A comparison of the experimental findings with the {\footnotesize AMPT} model presented in refs.43 and 44 for 2.76 TeV Pb-Pb collisions also suggests that {\footnotesize AMPT} too does not give a quantitative description of the data. However, in {\footnotesize AMPT} the resonances production at the hadronization phase due to hadronic re-scattering introduces additional correlation between particles, which, in turn, drives $\nu_{dyn}$ results toward negative values as the collision centrality increases, particularly for [$\pi$,p] combination the {\footnotesize AMPT}, contrary to the data, predicts negative values for semi-central and central collisions.\\
 
\noindent Shown in Fig.8 are dependence of $\nu_{dyn}[A,B]$ on the mean charged particle density for the three particle type pairs and three tunes of \hij at the three incident energies. These plots are obtained for the p$_{T}$ range, 0.2 $<$ p$_{T}$ $<$ 1.5 GeV/c. Similar plots for 0.2 $<$ p$_{T}$ $<$ 5.0 GeV/c are presented in Fig.9.
In order to examine the system size dependence, values of $\nu_{dyn}[A,B]$ for pp collisions events, simulated using {\footnotesize PYTHIA8} \cite{bib:68} at $\sqrt{s}$ = 2.76 and 5.02 TeV (100 $\times$ 10$^{6}$ events in each sample), are also shown in the figures. It is interesting to note from the figures that with increasing \dnch, the $\nu_{dyn}[A,B]$ values decreases first quickly, then slowly and finally tend to saturate for \dnch $\sim$ 100 and beyond. It may be of interest to note that data points corresponding to the {\footnotesize ALICE} experiment results [43] overlap the \hij data for each combination of particle pairs, except for very peripheral collisions where the experimental values for [p,K] and [$\pi$,p] are noticed to be somewhat smaller. A comparison of the results shown in Figs.8 and 9 indicates that $\nu_{dyn}$ values, if plotted against \dnch, essentially exhibit similar trends irrespective of the p$_{T}$ range considered. \\  


\noindent {\bf Conclusions} \\
\noindent On the basis of the findings of the present work, the following conclusions may be arrived at:
\begin{enumerate}
	\item For most-central collisions (0--5$\%$), \hij predicted values of $\nu_{dyn}$ for the three particle pairs follow the same trend as exhibited by the experimental data from {\footnotesize STAR} and {\footnotesize ALICE} experiments.	
	\item Values of $\nu_{dyn}$ for the three combinations of particle pairs for semi-central and peripheral collisions for Xe-Xe collisions are larger than those obtained for Pb-Pb collisions at $\sqrt{s}_{NN}$ = 2.76 and 5.02 TeV. This difference becomes larger with increasing collision centrality.
	\item A comparison of these findings with those reported by {\footnotesize ALICE} experiment for Pb-Pb collisions at $\sqrt{s}_{NN}$ = 2.76 TeV indicates that for [$\pi$,K] pair \hij under estimates, while for [p,K] and [$\pi$,p] pairs, the model overestimates the $\nu_{dyn}$ values. The observed difference in the \hij predicted and experimental values increases on moving from central to peripheral collisions. The observed lower values of $\nu_{dyn}[\pi,K]$ against centrality as compared to the experimental results suggests that there might be an anti-correlation between $\pi$ and K in the model or/and the multiplicity distributions of K and $\pi$ are broader. The higher values of $\nu_{dyn}$ for [p,K] and [$\pi$,p] combinations predicted by \hi, as compared to the experimental findings, may be due to rather weaker k-p and $\pi$-p correlations.	
	\item $\nu_{dyn}$ values for p$_{T}$ range 0.2 to 5.0 GeV/c, when scaled with charged particle density, are observed to acquire nearly constant values against collision centrality and beam energy. The effect of jet quenching and resonances are also observed to be absent. However, for a lower p$_{T}$ range a significant contribution from the resonance decays in the case of Pb-Pb collisions is observed for [$\pi$,p] and [p,K] pairs. The findings suggest too that the model predicted scaled values of $\nu_{dyn}[\pi,K]$ for Pb-Pb collisions are in close agreement with the experimental results. For [$\pi$,p] and [p,K] pairs the model predicts relatively higher values for all centrality classes in comparison to those observed in {\footnotesize ALICE} experiment.
	\item $\nu_{dyn}$ values for various collision systems, including (pp collisions) at {\footnotesize LHC} energies and for different tunes of \hi, when plotted against mean charged particle density give a smooth trend. With increasing \dnch, the values of $\nu_{dyn}$ first decreases upto \dnchh $\sim$ 100 and thereafter acquire saturation. The results reported based on {\footnotesize ALICE} data are also observed to be in accord with the model based findings.
\end{enumerate}

\newpage
\bibliographystyle{utphys}
	
\newpage
\begin{table}
	\centering
	\caption{Details of events simulated for the analysis.}\vspace{2mm}
	\begin{tabular}{c|c|c|c} \hline 
		Energy &  Type of   & Analysis    & No. of events\\
		(GeV)  & collision  &  Mode	      &	($\times 10^{6}$) \\  [2mm] \hline
		&    &                Res-off JQ-on    & 4.4   \\
		2760  &  Pb-Pb    &   Res-off JQ-off   & 3.6  \\  
		&     &               Res-on JQ-off    & 2.3  \\  \hline
		&     &               Res-off JQ-on    & 3.3   \\
		5020  &  Pb-Pb    &   Res-off JQ-off   & 2.5   \\ 
		&     &               Res-on JQ-off    & 2.3   \\ \hline
		&     &               Res-off JQ-on    & 2.9   \\
		5440  &  Xe-Xe    &   Res-off JQ-off   & 2.5  \\ 
		&     &               Res-on JQ-off    & 3.7  \\  \hline 
\end{tabular}
\end{table}

\newpage
\begin{table}[h]
	\centering	
	\caption{\footnotesize {Values of \npp and \dnch, in different centrality bins for three tunes of \hi, i) default, ii) Res-off JQ-on, iii) Res-on JQ-off for Pb-Pb collisions at $\sqrt{s}_{NN}$ = 2.76 TeV. The values within brackets are from the experiments, taken from ref.65$^{\ast}$. }}\vspace{4mm}
	\footnotesize
	\begin{tabular}{c|rr|rr|rr} \hline \hline
		Centrality&\multicolumn{2}{|c|}{\hij Res-off JQ-off}&\multicolumn{2}{|c|}{\hij Res-off JQ-on}& \multicolumn{2}{|c}{\hij Res-on JQ-off} \\  \cline{2-7}
		$(\%)$  & \npp & \dnchh &  \npp  & \dnchh & \npp & \dnchh \\ \hline
		0-5& 384.12$\pm$0.16&1891.4$\pm$0.64&  383.0$\pm$0.12&3609.66$\pm$3.88 &  383.48$\pm$0.16& 3086.72$\pm$2.57\\ 
		&(382.8$\pm$3.1)&(1601$\pm$60)& & &  &\\ \hline		
		5-10 &  326.94$\pm$0.15 & 1556.96$\pm$0.47 &  326.1$\pm$0.14 & 2885.69$\pm$2.97 &  325.49$\pm$0.14 & 2352.9$\pm$1.81\\ 
		& (329.7$\pm$4.6) & (1294$\pm$49) & & & &\\ \hline
		10-20 &  250.03$\pm$0.21 & 1198.73$\pm$0.42 &  249.8$\pm$0.21 & 2114.36 $\pm$2.70&  250.29$\pm$0.21 & 1727.05$\pm$ 0.97 \\ 
		& (260.5$\pm$4.4) & (966$\pm$37) & & & & \\ \hline
		20-30 &  170.60$\pm$0.18 & 841.22$\pm$0.31 &  170.9$\pm$0.17 & 1387.39$\pm$1.93&  170.64$\pm$0.17 & 1169.5$\pm$0.61 \\ 
		& (186.4$\pm$3.9) & (649$\pm$23) & & & &\\ \hline
		30-40 &  115.52$\pm$0.14 & 576.68$\pm$0.24 & 111.5$\pm$0.14 & 870.71$\pm$0.02 &  111.55$\pm$0.14 & 775.09$\pm$0.38 \\ 
		& (128.9$\pm$3.3) & (426$\pm$15) & & &  &\\ \hline
		40-50 &  68.57$\pm$0.12 & 381.47$\pm$0.18 & 68.9$\pm$0.11 & 519.95$\pm$0.92&  68.32$\pm$0.11 & 490.03$\pm$0.23 \\ 
		& (85.0$\pm$2.6) & (261$\pm$9) & & & & \\ \hline
		50-60 &  38.87$\pm$0.09 & 239.56$\pm$0.13 &  38.7$\pm$0.08 & 296.59$\pm$0.60 &  39.20$\pm$0.08& 293.2$\pm$0.13\\ 
		& (52.8$\pm$2.0) & (149$\pm$6) & & & &\\ \hline
		60-70 &  20.29$\pm$0.06 & 141.69$\pm$0.09 & 20.0$\pm$0.06 & 157.44$\pm$0.38&  20.21$\pm$0.06& 165.93$\pm$0.08 \\ 
		& (30.3$\pm$1.3) & (76$\pm$4) & & & &\\ \hline
		70-80 &  10.33$\pm$0.04 & 79.14$\pm$0.06 & 10.5$\pm$0.04 & 79.57$\pm$0.22&  10.44$\pm$0.04& 88.10$\pm$0.05 \\ 
		& (15.8$\pm$0.6) & (35$\pm$2) & & &  &\\ \hline

\multicolumn{7}{l}{$^{\ast}$Errors associated include systematic and statistical errors.} \\
	\end{tabular}
\end{table}

\newpage
\begin{table}
	\centering	
	\caption{\footnotesize {Values of \npp and \dnchh for Pb-Pb collisions at $\sqrt{s}_{NN}$ = 5.02 TeV. The same details are mentioned in Table~2. The values within brackets have been obtained from the data as reported in ref.66$^{\ast}$. }}\vspace{4mm}
	\footnotesize
		\begin{tabular}{c|rr|rr|rr} \hline \hline
		Centrality&\multicolumn{2}{|c|}{\hij Res-off JQ-off}&\multicolumn{2}{|c|}{\hij Res-off JQ-on}& \multicolumn{2}{|c}{\hij Res-on JQ-off} \\  \cline{2-7}
		$(\%)$  & \npp & \dnchh &  \npp  & \dnchh & \npp & \dnchh \\ \hline
		0-5   & 385.82$\pm$0.11   &  2338.57$\pm$0.5   	& 385.80$\pm$0.11  & 4463.09$\pm$1.52 & 385.88$\pm$0.16  & 3762.82$\pm$3.74\\ 
		& (385.0$\pm$2.5)   &  (1943$\pm$54)    & & & &\\ \hline		
		5-10   	& 331.04$\pm$0.14   &  1847.71$\pm$0.36   & 330.79$\pm$0.14  & 3344.38$\pm$0.94 & 330.57$\pm$0.14  & 2774.66$\pm$2.39 \\ 
		& (333.0$\pm$4.0)   &  (1586$\pm$46)     & & & &\\ \hline
		10-20   & 254.51$\pm$0.21   & 1344.25$\pm$0.29    & 254.04$\pm$0.21  & 2348.47$\pm$0.78 & 254.12$\pm$0.21  & 1942.26$\pm$1.29 \\ 
		& (263.0$\pm$4.0)   &  (1180$\pm$31)      & & & &\\ \hline
		20-30  	& 174.85$\pm$0.18  &  908.39$\pm$0.19     & 175.31$\pm$0.14  & 1489.35$\pm$0.50 & 174.89$\pm$0.17  & 1266.45$\pm$0.71 \\ 
		& (188.0$\pm$3.0)  &  (786$\pm$20)      & & & &\\ \hline
		30-40   & 116.56$\pm$0.15  &  615.52$\pm$0.14  	& 117.15$\pm$0.14  & 928.62$\pm$0.34 & 116.18$\pm$0.14  & 815.25$\pm$0.42 \\ 
		& (131.0$\pm$2.0)  &  (512$\pm$15)       & & & &\\ \hline
		40-50  	& 72.26$\pm$0.12  &  406.44$\pm$0.10     & 71.23$\pm$0.11  & 556.89$\pm$0.23 & 72.10$\pm$0.11  & 510.12$\pm$0.24 \\
		& (86.3$\pm$1.7)   &  (318$\pm$12)       & & & &\\ \hline
		50-60  	& 41.77$\pm$0.09  &  257.53$\pm$0.07      & 41.82$\pm$0.08  & 320.21$\pm$0.15 & 41.83$\pm$0.09  & 306.40$\pm$0.14\\ 
		& (53.6$\pm$1.2)  &  (183$\pm$8)           & & & &\\ \hline
		60-70   & 22.31$\pm$0.06  & 155.53$\pm$0.05  	& 22.09$\pm$0.06  & 175.42$\pm$0.09  & 21.98$\pm$0.06  & 174.66$\pm$0.08\\ 
		& (30.4$\pm$0.8)  &  (96.3$\pm$5.8)        & & & &\\ \hline
		70-80  	& 11.57$\pm$0.04  &  88.87$\pm$0.03  & 11.48$\pm$0.04 & 91.58$\pm$0.06 & 
		11.47$\pm$0.04  & 94.18$\pm$0.05 \\ 
		& (15.6$\pm$0.5)  &  (44.9$\pm$3.4)        & & & &\\ \hline

		\multicolumn{7}{l}{$^{\ast}$Errors associated include systematic and statistical errors.} \\
	\end{tabular}
\end{table}


\newpage
\begin{table}
	\centering	
	\caption{\footnotesize {Values of \npp and \dnchh for 5.44 TeV Xe-Xe collisions. The other details are as given in Table~2. Experimentally observed values, given within brackets, are taken from ref.67$^{\ast}$. }}\vspace{4mm}
	\footnotesize
	\begin{tabular}{c|rr|rr|rr} \hline \hline
		Centrality&\multicolumn{2}{|c|}{\hij Res-off JQ-off}&\multicolumn{2}{|c|}{\hij Res-off JQ-on}& \multicolumn{2}{|c}{\hij Res-on JQ-off} \\  \cline{2-7}
		$(\%)$  & \npp & \dnchh &  \npp  & \dnchh & \npp & \dnchh \\ \hline
		0-5    &  231.04$\pm$0.03   &  1518.05$\pm$0.33  & 231.31$\pm$0.04  & 2983.14$\pm$0.76 & 236.50$\pm$0.03  &  1787.88$\pm$0.33\\ 
		&  (236.0$\pm$2.0)   &  (1167$\pm$26)     &      	&       &           &  \\ \hline	
		5-10   & 202.44$\pm$0.04   &  1211.96$\pm$0.25  & 201.53$\pm$0.04  & 2288.45$\pm$0.54  & 208.43$\pm$0.03  &  1447.17$\pm$0.26  \\ 
		&(207.0$\pm$3.0)   &  (939$\pm$24)      & 	    &          &        &  \\ \hline	
		10-20   & 160.91$\pm$0.04   &  891.22$\pm$0.22  	& 159.98$\pm$0.04  & 1591.45$\pm$0.47 & 165.96$\pm$0.03  &  1059.88$\pm$0.23 \\ 
		&       (165.0$\pm$3.0)   &  (706$\pm$17)       & 	    &          &        &  \\ \hline	
		20-30   & 114.89$\pm$0.03  &  579.72$\pm$0.15  	& 114.26$\pm$0.03  & 956.1$\pm$0.31  &  119.15$\pm$0.03  &  688.21$\pm$0.16\\ 
		&   (118.0$\pm$4.0)  &  (478$\pm$11)       & 	        &          &        & \\ \hline	
		30-40     & 79.50$\pm$0.02  &  363.21$\pm$0.11 & 79.0$\pm$0.02  & 542.51$\pm$0.21 &  82.79$\pm$0.02  &  430.86$\pm$0.11\\ 
		& (82.2$\pm$3.9)  &  (315$\pm$8)       & 	    &          &        &  \\ \hline	
		40-50   & 52.69$\pm$0.02  &  214.87$\pm$0.08  & 52.29$\pm$0.02  & 287.15$\pm$0.13 &  55.15$\pm$0.02  &  255.61$\pm$0.08\\ 
		& (54.6$\pm$3.6)   &  (198$\pm$5)        &         	&       &           &  \\\hline	
		50-60   & 32.92$\pm$0.01  &  118.64$\pm$0.05  	& 32.76$\pm$0.01  & 142.38$\pm$0.07 &  34.77$\pm$0.01  &  141.99$\pm$0.05\\ 
		& (34.1$\pm$3.0)  &  (118$\pm$3)        & 	        &      &           & \\ \hline	
		60-70   & 19.28$\pm$0.01  &  60.98$\pm$0.03  	& 19.2$\pm$0.01  & 66.75$\pm$0.04 &     20.47$\pm$0.01  &  73.25$\pm$0.03\\ 
		& (19.7$\pm$2.1)  &  (64.7$\pm$2.0)     & 	        &     &            & \\\hline	
		70-80   & 10.54$\pm$0.01  &  28.88$\pm$0.02 	& 10.51$\pm$0.01  & 29.74$\pm$0.02 &   11.23$\pm$0.01  &  34.82$\pm$0.02\\ 
		& (10.5$\pm$1.1)  &  (32.0$\pm$1.3)     & 	              &                & &  \\ \hline	
		
		\multicolumn{7}{l}{$^{\ast}$Errors associated include systematic and statistical errors.} \\
	\end{tabular}
\end{table}



\newpage
\begin{figure}[htb!]
	\begin{center}~
		\includegraphics*[width=16.0cm,height=16.0cm,keepaspectratio=true]{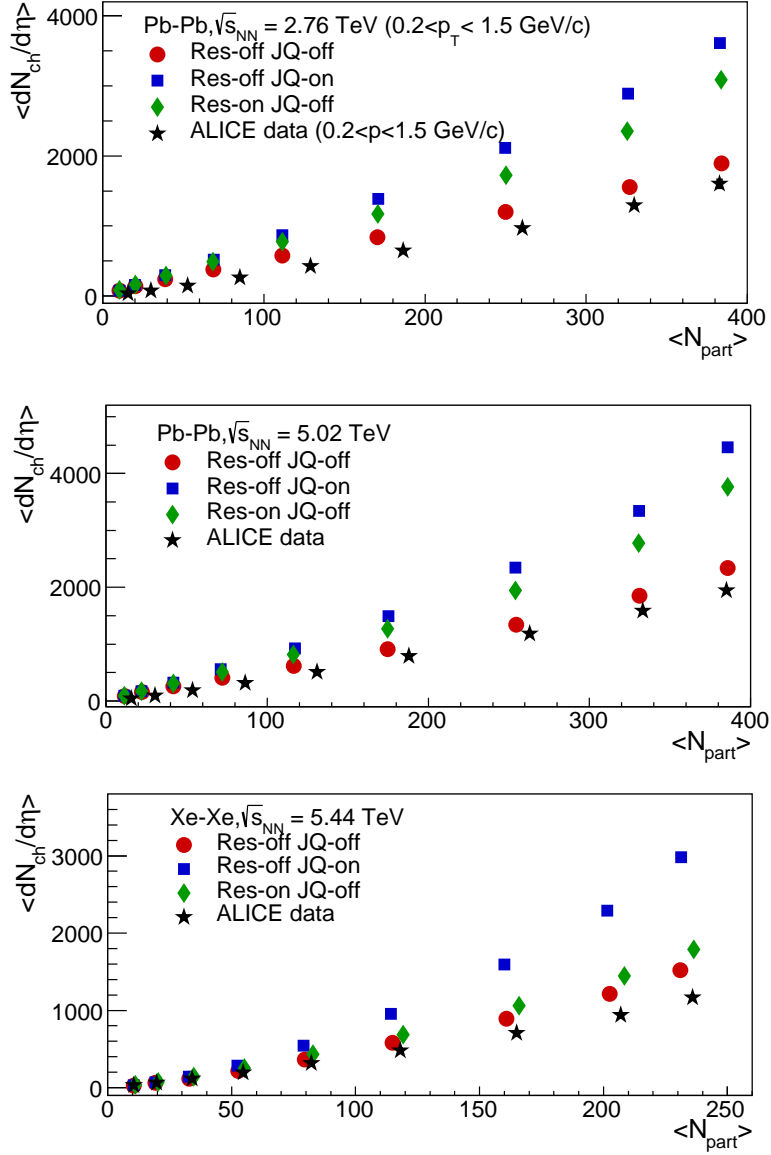}
		\caption{Dependence of charged particle density on mean number of participating nucleons for Pb-Pb collisions at $\sqrt{s}_{NN}$ = 2.76 and 5.02 TeV and Xe-Xe collisions at $\sqrt{s}_{NN}$ = 5.44 TeV. The values reported by {\footnotesize ALICE} collaboration are taken from refs.65-67.  }
		
		\label{fig:dnchdetaVsNpart}
	\end{center}
\end{figure}
\begin{figure}[htb!]
	\begin{center}~
		\includegraphics*[width=16.0cm,height=16.0cm,keepaspectratio=true]{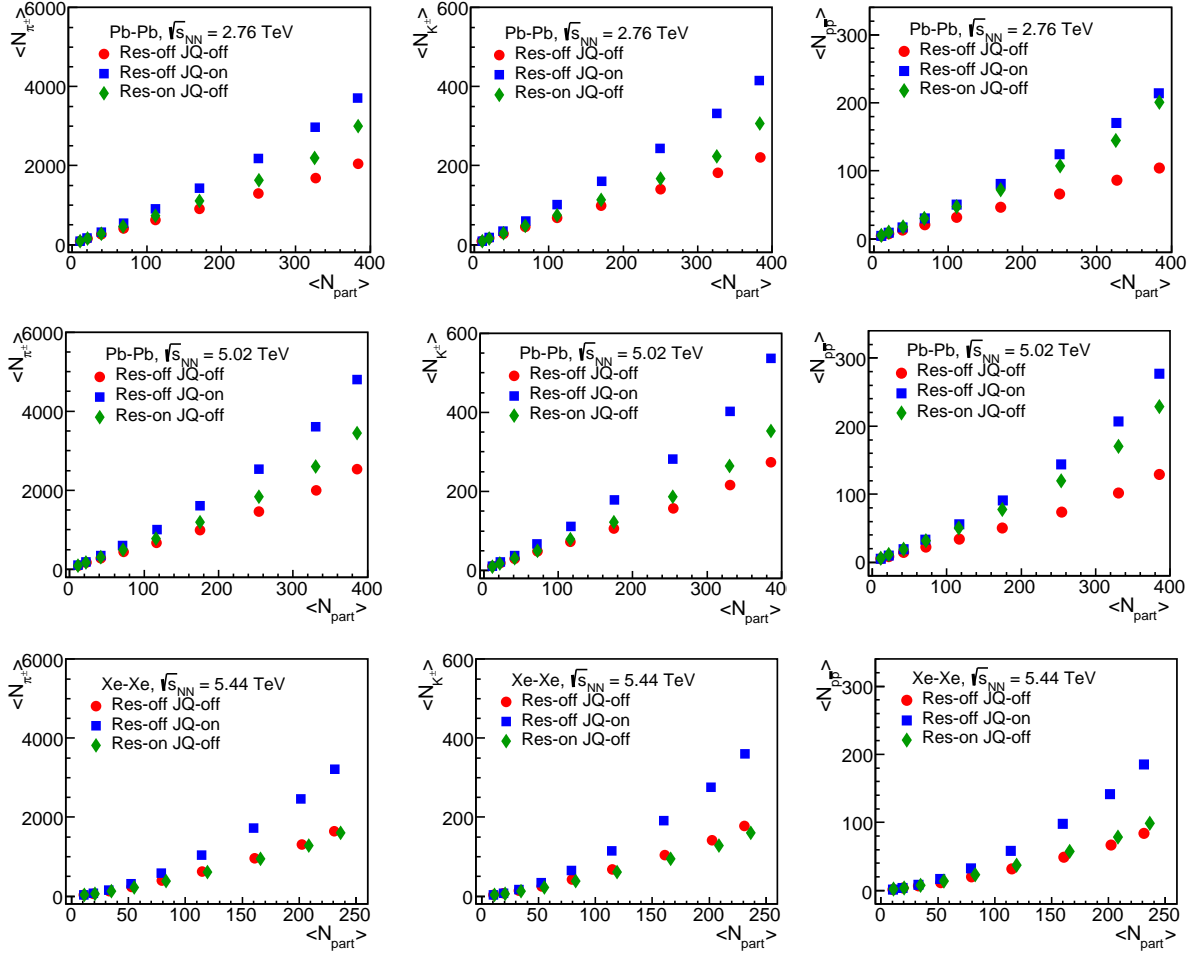}
		
		\caption{Variations of $\langle$N$_{ch}$$\rangle$ with $\langle$N$_{part}$$\rangle$ for $\pi^{\pm}$, $K^{\pm}$ and $p\bar{p}$ in Pb-Pb and Xe-Xe collisions at $\sqrt{s}_{NN}$ = 2.76, 5.02 TeV and 5.44 TeV. The transverse momentum range is 0.2$<$p$_{T}$$<$1.5 GeV/c. }
		
		\label{fig:DiffNch}
	\end{center}
\end{figure}

\newpage
\begin{figure}[htb!]
	\begin{center}~
		\includegraphics*[width=16.0cm,height=16.0cm,keepaspectratio=true]{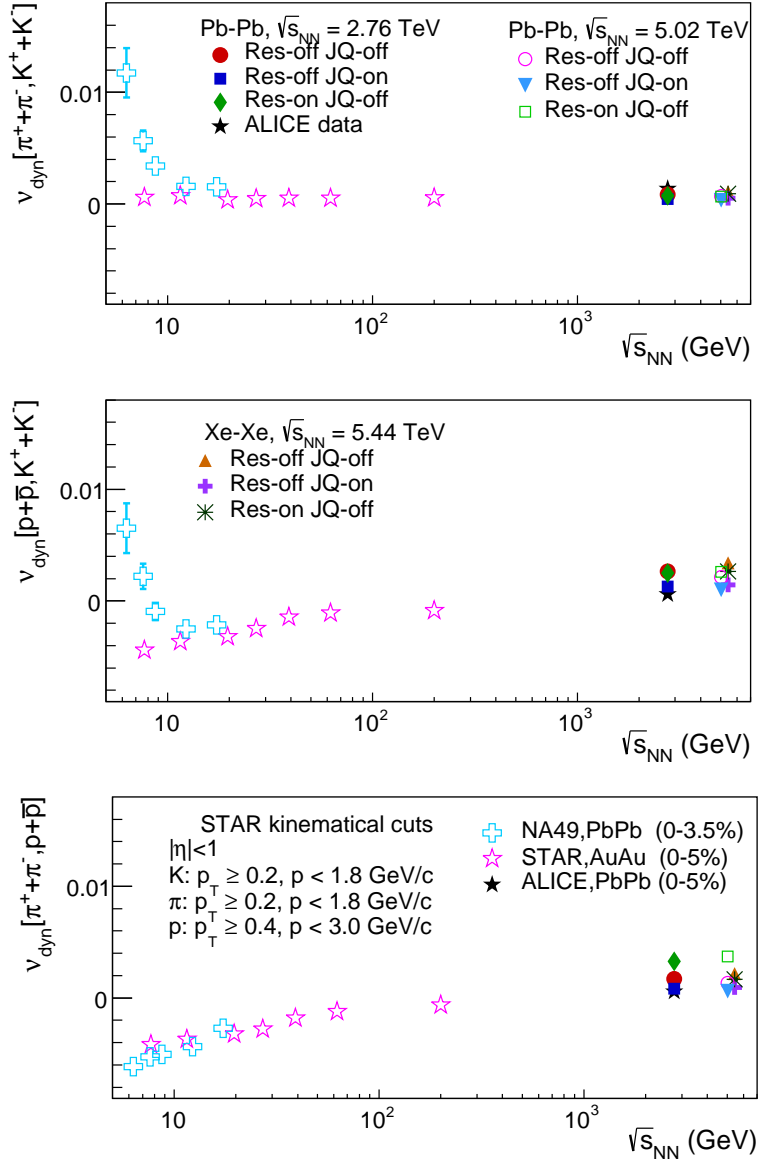}
		
		\caption{Energy dependence of $\nu_{dyn}$ for $[\pi,K]$, $[p,K]$, $[\pi,p]$ for \hij default, Res-off JQ-on and Res-on JQ-off for 0--5$\%$ central events. Values corresponding to {\footnotesize STAR}, {\footnotesize NA49} and {\footnotesize ALICE} experiments are taken from refs.40,41,43.  }
		
		\label{fig:EnergyPlot}
	\end{center}
\end{figure}

\newpage
\begin{figure}[htb!]
	\begin{center}~
		\includegraphics*[width=16.0cm,height=16.0cm,keepaspectratio=true]{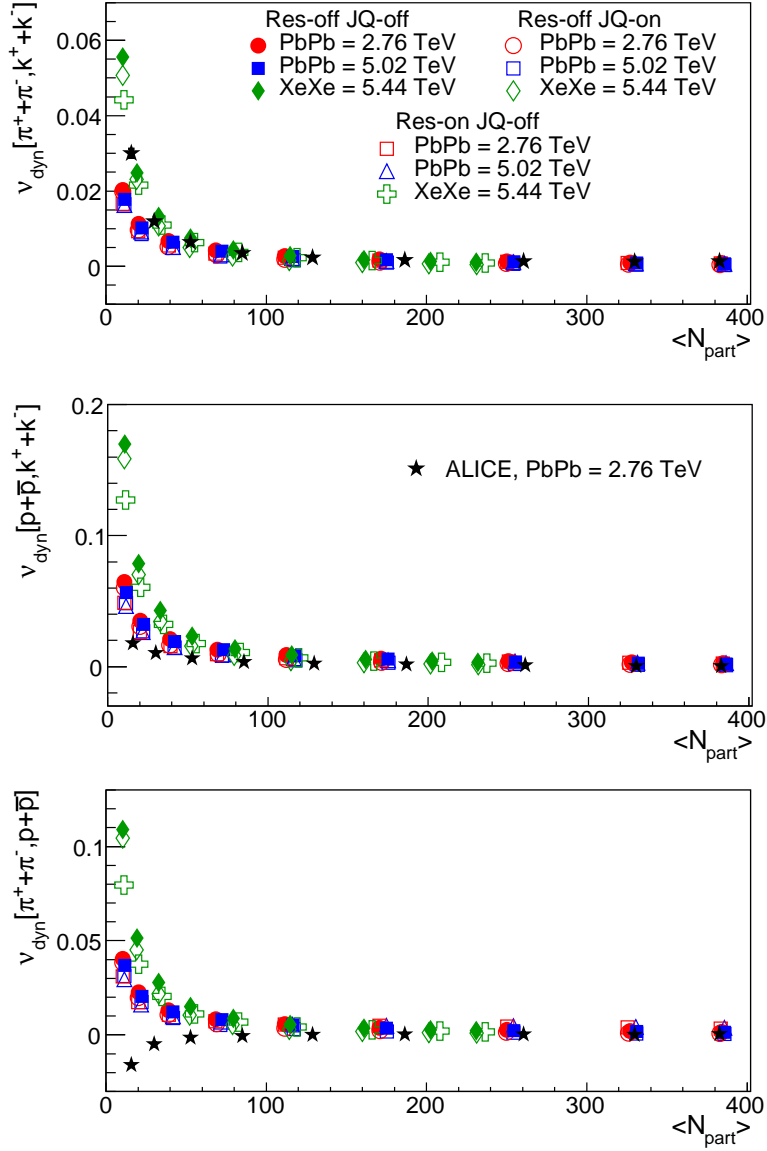}
		
		\caption{$\langle$N$_{part}$$\rangle$ dependence of $\nu_{dyn}[\pi,k]$, $\nu_{dyn}[p,K]$ and $\nu_{dyn}[\pi,p]$ for 0.2$<$p$_{T}$$<$1.5 GeV/c. Experimental values for Pb-Pb collisions at $\sqrt{s}_{NN}$ = 2.76 TeV, shown in the figure, are taken from ref.43. }
		
		\label{fig:NudynVsCent}
	\end{center}
\end{figure} 

\newpage
\begin{figure}[htb!]
	\begin{center}~
		\includegraphics*[width=16.0cm,height=16.0cm,keepaspectratio=true]{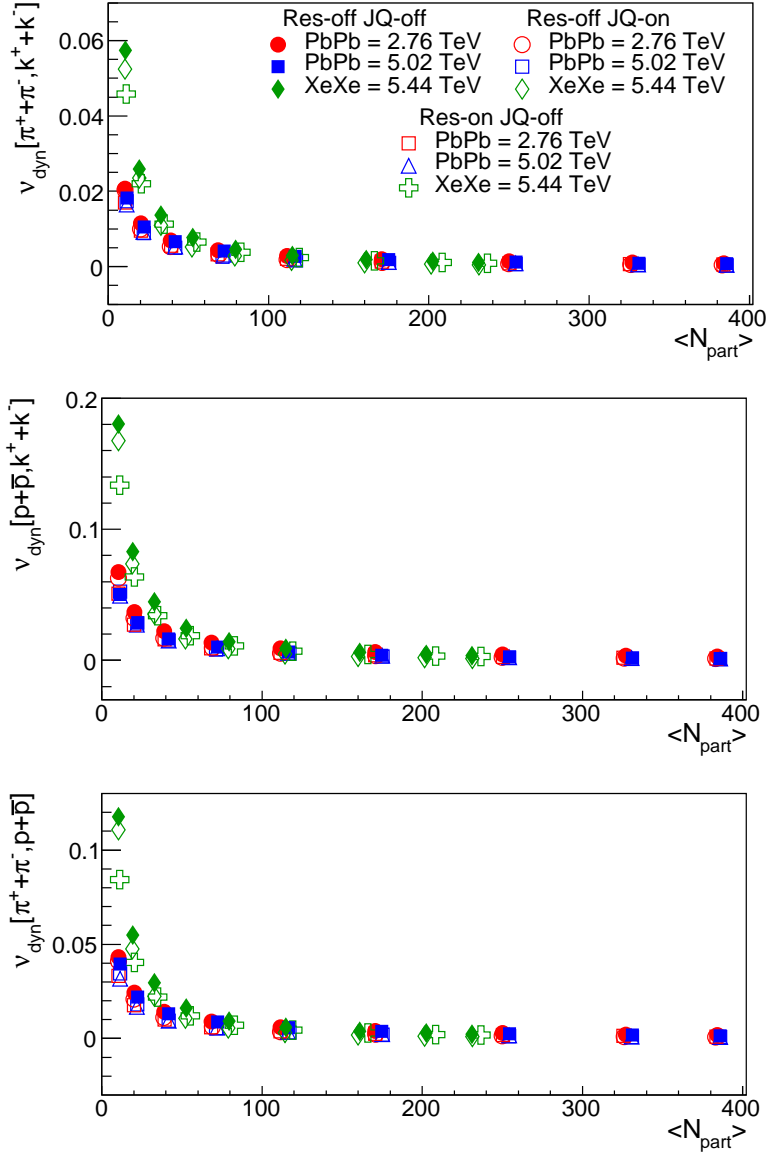}
		
		\caption{ Variations of $\nu_{dyn}[\pi,k]$, $\nu_{dyn}[p,K]$ and $\nu_{dyn}[\pi,p]$ with $\langle$N$_{part}$$\rangle$ for 0.2$<$p$_{T}$$<$5.0 GeV/c.}
		
		\label{fig:NudynVsCent2}
	\end{center}
\end{figure}

\newpage
\begin{figure}[htb!]
	\begin{center}~
		\includegraphics*[width=16.0cm,height=16.0cm,keepaspectratio=true]{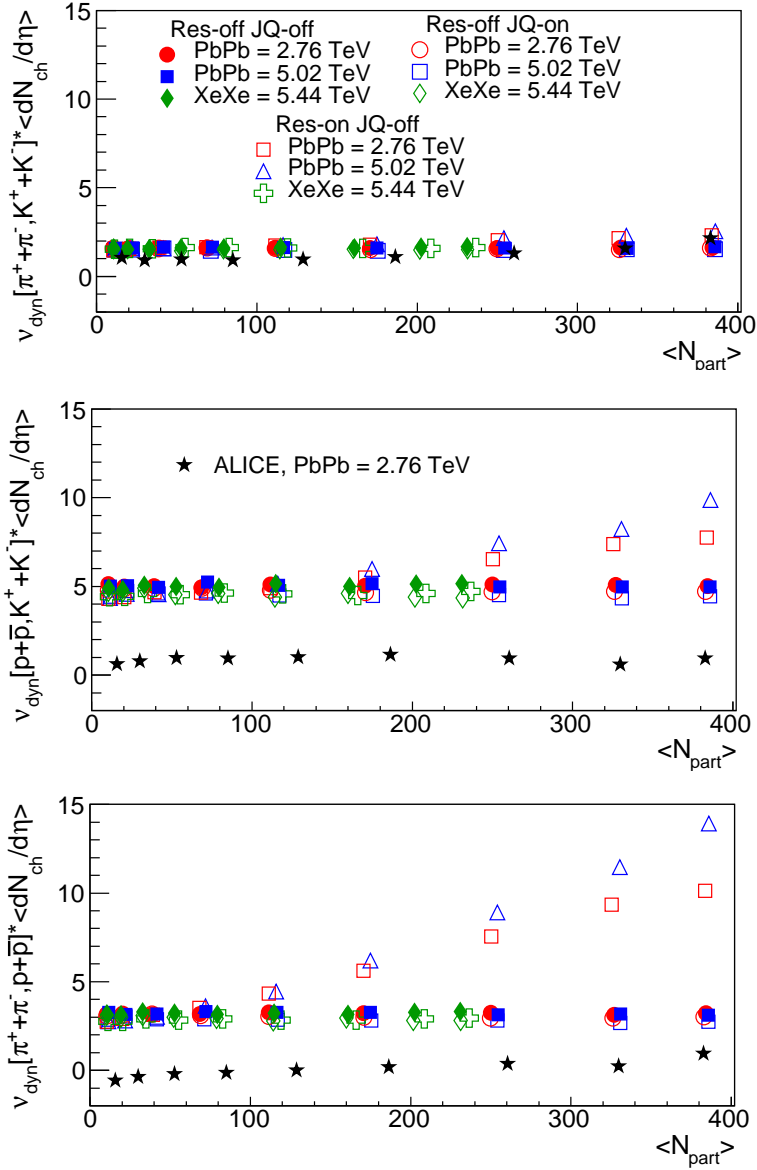}
		
		\caption{$\nu_{dyn}$ scaled by \dnchh vs $\langle$N$_{part}$$\rangle$ for \hij events in p$_{T}$ range, 0.2$<$p$_{T}$$<$1.5 GeV/c. {\footnotesize ALICE} experimental values, shown in the figure, are taken from ref.43.}
		
		\label{fig:ScaledNudyn}
	\end{center}
\end{figure}

\newpage
\begin{figure}[htb!]
	\begin{center}~
		\includegraphics*[width=16.0cm,height=16.0cm,keepaspectratio=true]{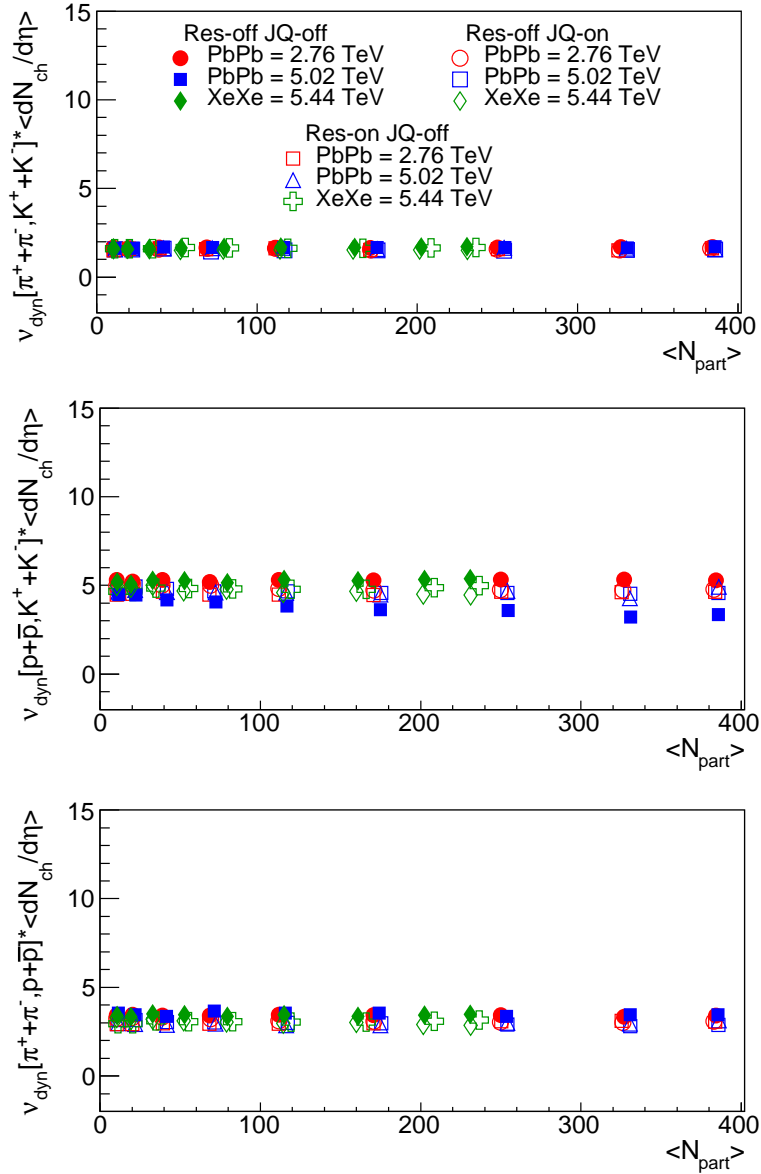}
		
		\caption{$\nu_{dyn}$ scaled by \dnchh vs $\langle$N$_{part}$$\rangle$ in the p$_{T}$ range, 0.2$<$p$_{T}$$<$5.0 GeV/c.}
		
		\label{fig:ScaledNudyn_2}
	\end{center}
\end{figure}

\newpage
\begin{figure}[htb!]
	\begin{center}~
		\includegraphics*[width=16.0cm,height=16.0cm,keepaspectratio=true]{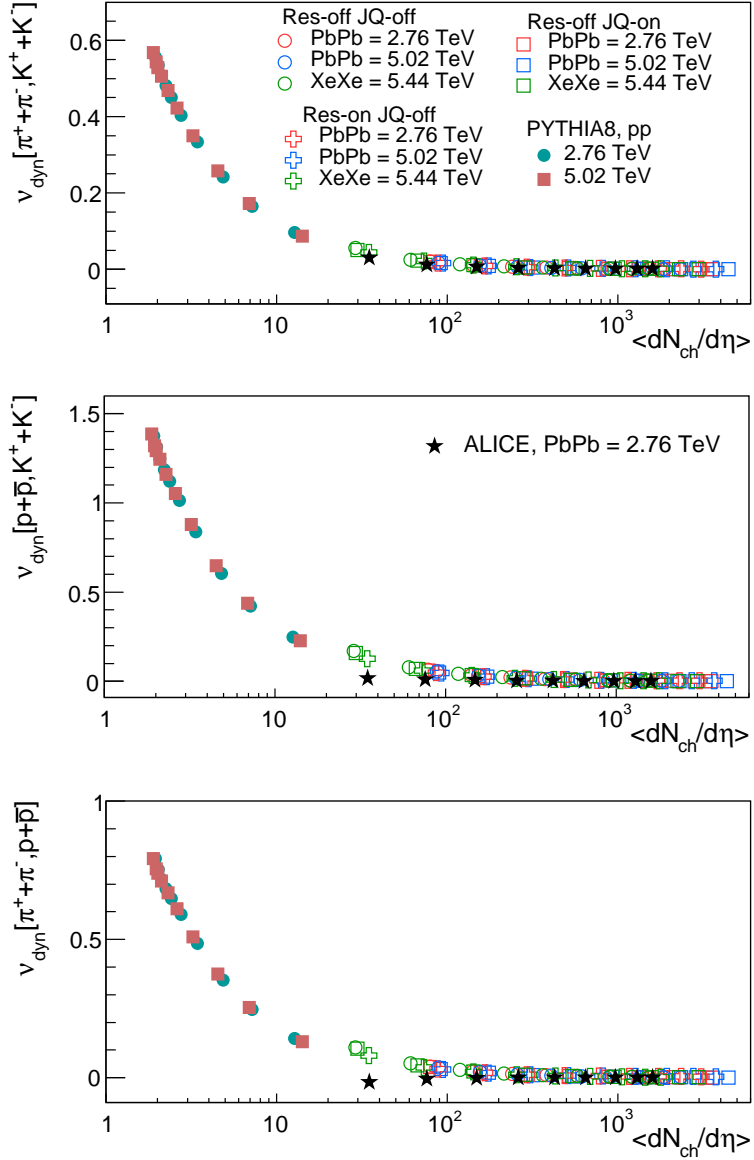}
		
		\caption{\dnchh dependence of $\nu_{dyn}[\pi,k]$, $\nu_{dyn}[p,K]$ and $\nu_{dyn}[\pi,p]$ for pp and Pb-Pb collisions at $\sqrt{s}_{NN}$ = 2.76, 5.02 TeV and Xe-Xe collisions at $\sqrt{s}_{NN}$ = 5.44 TeV in the p$_{T}$ range, 0.2$<$p$_{T}$$<$1.5 GeV/c. Experimental results for Pb-Pb collisions at $\sqrt{s}_{NN}$ = 2.76 TeV, shown in the figure are taken from ref.43.}
		
		\label{fig:DiffMom1}
	\end{center}
\end{figure}

\newpage
\begin{figure}[htb!]
	\begin{center}~
		\includegraphics*[width=16.0cm,height=16.0cm,keepaspectratio=true]{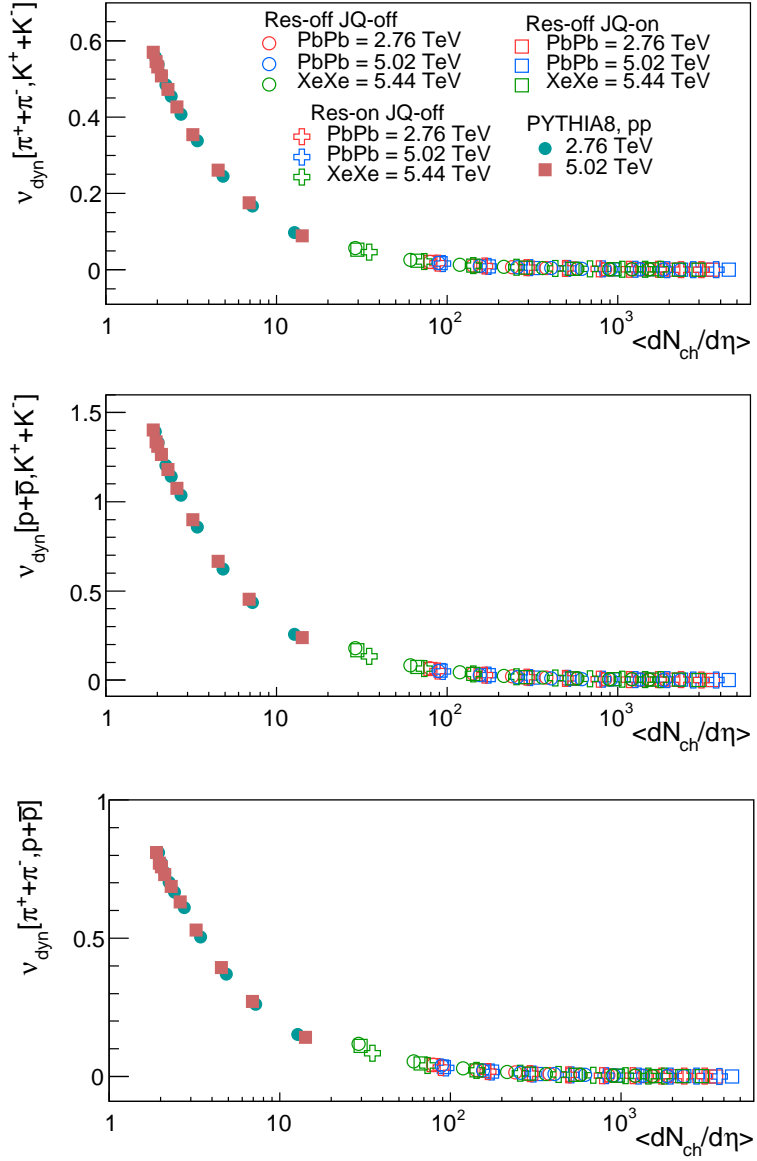}
		
		\caption{\dnchh dependence of $\nu_{dyn}[\pi,k]$, $\nu_{dyn}[p,K]$ and $\nu_{dyn}[\pi,p]$ for various collision systems at $\sqrt{s}_{NN}$ = 2.76, 5.02 and 5.44 TeV in the p$_{T}$ range, 0.2$<$p$_{T}$$<$5.0 GeV/c.}
		
		\label{fig:DiffMom2}
	\end{center}
\end{figure}

\end{document}